\documentclass{article} 
\usepackage[german,english]{babel}
\usepackage{fontenc}
\usepackage{amssymb}
\usepackage{subfigure}
\usepackage{graphicx}
\usepackage{color}
\usepackage{units}
\usepackage{hyperref}
\usepackage[table]{xcolor}

\usepackage[version=3]{mhchem}
\hypersetup{colorlinks,%
           citecolor=blue,%
          filecolor=blue,%
           linkcolor=blue,%
           urlcolor=blue,%
       pdfborder=0 0 0}
\usepackage[version=3]{mhchem} 
\usepackage{chemfig} 
\usepackage[utf8]{inputenc}
\usepackage{csquotes} 

\usepackage{placeins}
\newcommand{\nbar}{\text{$\bar{n}$} }

\begin{document}

\title{Simulation of weak polyelectrolytes: \\A comparison between the constant pH and the reaction ensemble method}
\author{Jonas Landsgesell\thanks{Institute for Computational Physics, University of Stuttgart, D-70569 Stuttgart, Germany} \and Christian Holm \and Jens Smiatek }

\abstract{ 
The reaction ensemble and the constant pH method are well-known chemical equilibrium approaches to simulate protonation and deprotonation reactions in classical molecular dynamics and Monte Carlo simulations.
In this article, we {show similarity} between both methods {under certain conditions}. We perform molecular dynamics simulations of a weak polyelectrolyte  in order to compare the titration curves obtained by both approaches. Our findings reveal a good agreement between the methods when the reaction ensemble is used to sweep the reaction constant. Pronounced differences between the reaction ensemble and the constant pH method can be observed for stronger acids and bases in terms of adaptive pH values. These deviations are due to the presence of explicit protons in the reaction ensemble method which induce a screening of electrostatic interactions between the charged titrable groups of the polyelectrolyte. The outcomes of our simulation hint to a better applicability of the reaction ensemble method for systems in confined geometries and titrable groups in polyelectrolytes with different pK$_\text{a}$ values. 
} 
\maketitle

\section{Introduction}
Weak polyelectrolytes, like polyacrylic acid or most proteins \cite{richtering06a} have titrable groups which can be either in a protonated or a deprotonated state depending on the pH value of the solution  \cite{berg02a}. The influence of the pH value gives rise to phenomena like protonation-configuration \cite{castelnovo00a,shi12a}, or charge regulation effects \cite{lund05a,lund13a} as they are known for weak polyelectrolytes like proteins.
A protonation/deprotonation reaction of a titrable group in a weak polyelectrolyte can be written as
\begin{center}
\ce{HA <=> A^- + H^{+}}\par 
\end{center}
where $\rm{HA}$ denotes the protonated form of the titrable group, $\rm{A^-}$ the deprotonated form and $\rm{H^+}$ the dissociated proton. The presence of water as a proton acceptor or proton donor is omitted for the sake of simplicity.
Depending on the pH value defined by $\rm{pH=-\log_{10}(c(H^+)/(\text{mol/L)})}$, the degree of association can be calculated via
\begin{equation}
\label{eq:nbar}
\nbar=\frac{N_\textrm{HA}}{N_0}
\end{equation}
with the number of associated titrable groups $N_\textrm{HA}$ divided by the total number of titrable groups $N_0=N_\textrm{HA}+N_\textrm{A$^-$}$, where $N_\textrm{A$^-$}$ denotes the number of deprotonated units \cite{harnung12a,butler98a}.
The equilibrium concentration of each species is steered by an apparent reaction constant in accordance to the law of mass action
\begin{equation}
\label{eq:law of mass action}
K_a=\frac{c(\textrm{A}^-)c(\textrm{H$^+$})}{c(\textrm{HA})}
\end{equation}
where $c(\cdot)$ denotes the concentration of the individual species \cite{atkins10a}. More conveniently, the logarithmic reaction constant is defined by 
$\rm{pK_a =-\log_{10}(K_a/(\text{mol/L)})}$, where typical values for $\rm{pK_a}$ vary between $\rm{pK_a =2-10}$ for common weak polyelectrolytes\,\cite{mueller13a}.  \\
The simulation of weak polyelectrolytes in classical molecular dynamics (MD) or Monte Carlo (MC) simulations is a challenging task. Over the last decades,  the constant pH and the reaction ensemble method were the most frequently used algorithms to perform these simulations. The constant pH method \cite{reed1992monte} implements a constant and global pH value as an input parameter which balances the probability of protonation and deprotonation reactions. The method was originally developed to simulate linear polyelectrolytes in presence and absence of excess salt and counterions \cite{reed1992monte,ullner96a,ullner96b,ullner2000simulations} and reveals a good agreement with analytical results \cite{jonsson1996titrating}. More effort was additionally spent on the study of different solvent conditions and their influence on polyelectrolyte conformations  \cite{uyaver2009effect,carnal2011chain} and the properties of polyampholytes \cite{nair2014conformational,lund05a}. Moreover, the constant pH method can also be employed in slightly modified versions \cite{mongan2004constant} in order to analyze shifts in the apparent reaction constant for proteins.\\
A different approach was introduced by the reaction ensemble (RE) method which also provides the possibility to model \textit{arbitrary} chemical reactions in classical simulations \cite{heath08a,smith94a,johnson1994reactive}.
The reaction ensemble method was used to study chemical and phase equilibria in different systems and to investigate the influence of high pressure and high temperature on chemical reactions \cite{heath08a}. Furthermore, it can be applied for acidic molecules in confined geometries, for the study of interface effects \cite{heath08a} and for the simulation of acid-base reactions in weak polyelectrolyte systems \cite{panagiotopoulos2009charge,uhlik2014modeling,uhlik2016charge,landsgesell2016}.
In contrast to the constant pH method, the reaction ensemble method considers an adaptive pH value which significantly differs from the global interpretation of the predefined implicit pH value in the constant pH method.\\
In this article, we elucidate the main properties of the constant pH and the reaction ensemble method. Moreover, we also demonstrate {similarity} between the reaction ensemble method and  {the constant pH method under certain conditions.} Furthermore,  a new interpretation of the reaction ensemble method is proposed in order to reproduce a real titration experiment. Our MD simulations verify that the explicit presence of protons in the reaction ensemble method induces electrostatic screening effects between the charged titrable groups of a polyelectrolyte in contrast to the constant pH simulations. Hence,  differences in the results obtained by both methods can be assigned to the different implementations of predefined and adaptive pH values.\\
The article is organized as follows. In the next section, we discuss the main properties of the reaction ensemble and the constant pH method.  Furthermore, we present a novel interpretation of the reaction ensemble method which is useful for the study of titration curves. In section 3, we show {similarity} between the constant pH and the reaction ensemble method {under certain conditions}. After the presentation of the numerical details in section 4, we compare the titration curves obtained by both methods for weak polyelectrolytes in coarse-grained MD simulations. We briefly conclude and summarize in the last section.  

\section{Properties of the constant pH and the reaction ensemble method}
The reaction ensemble and the constant pH method rely on Monte Carlo techniques and can be implemented in terms of the  Metropolis-Hastings algorithm \cite{hastings1970monte,frenkel2001understanding}. 
The Metropolis-Hastings algorithm is an importance sampling technique which fosters the population of states with a certain probability, for example the canonical probability $p_i= \exp(-\beta E_i)/Z$  with the energy of a state $i$, the inverse thermal energy $\beta=1/k_BT$ with the Boltzmann constant $k_B$, {the temperature $T$} and the canonical partition sum $Z$. A detailed balance condition between two states $r$ and $l$ is defined by 
\begin{equation}
p(r)t(l|r)=p(l)t(r|l),
\label{eq:detailed balance}
\end{equation}
with the probability $p$ and the associated transition probability $t$ between the two states. The transition probability itself can be calculated by the definition of a proposal probability $g$ and an acceptance probability $\text{acc}$ according to
\begin{equation}
t(r|l)=g(r|l)\text{acc}(r|l)
\end{equation} 
which can be inserted into Eqn.~\eqref{eq:detailed balance} to yield
\begin{equation}
\frac{\text{acc}(l|r)}{\text{acc}(r|l)}=\frac{p(l)g(r|l)}{p(r)g(l|r)}.
\end{equation}
A standard choice \cite{frenkel2001understanding} for the acceptance probability $\text{acc}(l|r)$ reads
\begin{equation}
\text{acc}(l|r)=\min\left(1,\frac{p(l)g(r|l)}{p(r)g(l|r)}\right),
\end{equation}
which fulfills the requirements of the detailed balance condition (Eqn.~\eqref{eq:detailed balance}). In the original Metropolis algorithm \cite{metropolis1953equation}, the proposal probability is symmetric \mbox{($g(l|r)=g(r|l)$)} and therefore the acceptance probability simplifies to
\begin{equation}
\text{acc}(l|r)=\min\left(1,\frac{p(l)}{p(r)}\right)
\end{equation}
for arbitrary {choices} of $g$. Specific expressions of $g$ are mostly intended to achieve an efficient and effective sampling of the phase space. 

\subsection{The reaction ensemble method}
In presence of chemical equilibrium, the reaction ensemble method proposes changes in the particle numbers of the reacting species by forward (deprotonation) and backward (protonation) reactions \cite{smith94a,heath08a,johnson1994reactive}. As outlined by Turner {\em et al.} \cite{heath08a}, the definition of the reaction ensemble with fluctuating particle numbers can be derived from the grand canonical ensemble via 
the separation of the kinetic and the configurational canonical partition sum.\\
In general, a chemical reaction can be written as 
\begin{equation}
\sum_{i=1}^z\nu_is_i=0
\end{equation}
for $z$ chemical species of type $s_i$ with stoichiometric coefficients $\nu_i$ \cite{atkins10a}.
The acceptance probability in the reaction ensemble for an arbitrary forward reaction from state $r$ to $l$ is defined as  
\begin{equation}
\label{eq:Metropolis_RE}
\text{acc}^{\text{RE}}(l|r)=\min\bigg\{1,(\beta P^0 V)^{\overline{\nu}\xi} K^{\xi} \prod_{i=1}^z\left[ \frac{N_i^0!}{(N_i^0+\xi \nu_i)!}\right]\exp(-\beta \Delta E_\text{pot})\bigg\},
\end{equation}
where $N_i$ is the number of particles after a reaction, $N_i^0$ the number of particles prior to a reaction and $\xi$ the \enquote{extent} of the reaction which is selected randomly with $\xi \pm 1$ \cite{heath08a}. A deprotonation (forward) reaction is defined by $\xi=+1$ and a protonation (backward) reaction by $\xi=-1$. Additional parameters are the dimensionless reaction constant $K$ for each titrable group, which is proportional to the apparent reaction constant in Eqn.~\eqref{eq:law of mass action}, the standard pressure $P^0$, the potential energy difference with $\Delta E_\text{pot}=E_\text{pot,r}-E_\text{pot,l}$, the volume of the system $V$ and the total change in the number of particles $\overline{\nu}=\sum_{i} \nu_i$.  The corresponding protonation and deprotonation reactions are usually performed after a predefined number of MD simulation steps with constant particle numbers \cite{heath08a}.
 
\subsection{Operation modes of the reaction ensemble method}
\label{sec:mode}
Two different options on how to apply the reaction ensemble method can be defined. We coin the first operation mode \enquote{sweeping the reaction constant} and our proposed second operation mode \enquote{real titration}. \\
The \enquote{sweeping} operation mode, which was for example applied in  Refs.~\cite{panagiotopoulos2009charge,uhlik2014modeling,uhlik2016charge,landsgesell2016} considers multiple independent simulation runs in which the individual titrable units are characterized by  various dimensionless input reaction constants $K=K_a /(\beta P_0)^{\overline{\nu}}$  \cite{landsgesell2016}. This definition has a direct chemical interpretation: in each {independent} simulation, titrable units with a {fixed} and arbitrary reaction constant $K$ are {inserted into \textit{initially} neutral water solution}. {After  equilibration of the reaction ensemble, a certain pH value in the simulation box is adjusted, which typically deviates from the neutral pH value. The corresponding pH value can be regarded as the equilibrium pH value for the given choice of the reaction constant $K$ and the given surrounding. In the following, we denote this pH value as the \enquote{eigen pH value}.} {By choosing} various values of $K$, the resulting degrees of association {as well as the corresponding (eigen) pH values can be obtained and therefore also the titration curves as a function of the pK$_a$-pH value.}\\
{
As a second operation mode, we propose to directly imitate a \enquote{titration} experiment \cite{atkins10a}. In a real titration experiment, a certain substance is titrated by injecting a strong acid or a strong base into the system. It is important to note, that the pH value changes in contrast to \textit{the reaction constant, which is a fixed substance property}. Therefore, in the real titration mode that we propose, we fix the intrinsic reaction constant and add certain amounts of a strong acid ({\em{e.~g.~}} HCl with species \ce{H+} and \ce{Cl-}) or a strong base ({\em{e.~g.~}} NaOH {with species \ce{Na+} and \ce{OH-}}) to the system. Due to their chemical properties \cite{atkins10a}, strong acids and bases reveal a very high dissociation constant and therefore remain dissociated even at extremely high or low pH values, respectively. Moreover, also the autoprotolysis reaction of water \ce{2 H2O <=> H_3O+ +OH-} with an apparent reaction constant of $K_w=10^{-14} \text{mol}^2/\text{L}^2$ is explicitly taken into account. Therefore, the protons and hydroxide ions can react to neutral water molecules which are neglected in our implicit solvent approach. It has to be noticed that hydroxide ions and protons differ from ordinary counterions (\ce{Na+} and \ce{Cl-}) only due to their properties in our algorithm, which enables them to participate in chemical reactions (autoprotolysis and deprotonation/protonation reactions of the polyelectrolyte), whereas all other properties are identical.\\ 
In order to approach higher or lower pH values than the eigen pH value of the weak acid, a strong base or a strong acid, respectively, are injected into the system.} 
Based on {the autoprotolysis reaction of water}, the concentration of the deprotonated or protonated titrable units of the weak polyacid directly adapts to the concentration of the excess H$^+$ or OH$^-$ species in the solution. In fact, this operational mode resembles an experimental titration procedure due to the explicit presence of protons or hydroxide ions and cannot be imitated by the constant pH method {without further effort}.
\subsection{The constant pH method}
\label{sec: similarity}
In the constant pH method, the protonation or deprotonation probability for a titrable group is determined after two steps. First, a random titrable group is chosen. If it is of type A$^-$ or HA, the group or particle, respectively and the corresponding properties are exchanged. Thus, dissociated protons are randomly placed in the simulation box and charge neutrality is fulfilled. The trial move is accepted with a probability
\begin{equation}
\text{acc}(l|r) =\min \left(1,\exp \left[-\beta \left( \Delta E_\text{pot} +\bigg(\pm (\ln(10)/\beta) \left(\text{pH}_{\text{in}}-\text{pK}_{a}\right)\bigg)\right)\right]\right)
\label{eq:acc const pH}
\end{equation}
where $\Delta E_\text{pot}$ is the potential energy change due to the exchange of chemical species, pH$_\text{in}$ is an input parameter which determines the implicit pH value of the solution and $\rm{pK_{a}}$ is the negative common logarithm of the apparent reaction constant, which is also a simulation parameter with a predefined value.  The expression $\pm (1/\beta)\left(\text{pH}_\text{in}-\text{pK}_{a}\right)$ can be interpreted as a change of the chemical potential \cite{reed1992monte}. A negative prefactor defines a deprotonation reaction (diss) and a positive prefactor, {\em vice versa} assigns a protonation reaction (ass).
Moreover, it has to be noted  that the proposal probability for a protonation or deprotonation reaction  in the constant pH method is asymmetric \cite{reed1992monte,ullner1996monte,ullner1996montesalt,jonsson1996titrating,uyaver2009effect,carnal2011chain,ullner2000simulations,nair2014conformational}. This can be shown by comparing  the proposal probability for a deprotonation reaction which reads $g(\text{diss}|\text{ass})=N_\text{HA}/N_0$, with the proposal probability for a protonation reaction defined as $g(\text{ass}|\text{diss})=N_{\text{A}^-}/N_0$, which implies $g(\text{ass}|\text{diss})\neq g(\text{diss}|\text{ass})$.

\section{The constant pH method with a symmetric proposal probability}
{As we have discussed above, the constant pH method relies on asymmetric proposal probabilities. In order to show {similarity} between the constant pH and the reaction ensemble method under certain conditions, we  develop an expression for the constant pH method with a symmetric proposal probability which is then compared to the reaction ensemble method.}
Noteworthy, the acceptance probability can be evaluated by the original Metropolis algorithm \cite{metropolis1953equation} with symmetric proposal probabilities per definition.  
In order to follow this approach, we use an expression for the partition sum of the constant pH ensemble which was proposed in Ref.~\cite{reed1992monte} and reads
\begin{equation}
Z_\text{pH}= \sum_{\nbar} {N_0 \choose (1-\nbar) N_0} x^{N_0 (1-\nbar)} \sum_{i(\nbar)} \exp(-\beta E_\text{pot,i}),
\label{eq: partition function constant pH}
\end{equation}
as a sum over all degrees of association and over all corresponding configurational microstates $i$ of the system.
The individual probability for a microstate with a certain degree of association reads
\begin{align}
p(\nbar,E_\text{pot, i})={N_0 \choose (1-\nbar) N_0} x^{N_0 (1-\nbar)} \exp(-\beta E_\text{pot,i})
\label{eq:const pH probability of state}
\end{align}
with $x=10^{\text{pH}_\text{in}-\text{pK}_{a}}$ and predefined and fixed values for  $\text{pH}_\text{in}$ and $\text{pK}_{a}$.
A deprotonation step for a single titrable group can be expressed by a change of the degree of association $\Delta \nbar$ in order to describe the transition from $(\nbar,E_\text{pot,ass})$ to $(\nbar-\Delta \nbar, E_\text{pot,diss})$.
Thus, the Metropolis acceptance probability \cite{metropolis1953equation} for this Monte Carlo move reads
\begin{equation}
\label{eq:tilde_acc}
\tilde{\text{acc}}(\text{diss}|\text{ass})=\min \left(1,\frac{p(\nbar-\Delta \nbar, E_\text{pot,diss})}{p(\nbar,E_\text{pot,ass})}\right)
\end{equation}
which yields 
\begin{equation}
\tilde{\text{acc}}(\text{diss}|\text{ass})=\min \left(1,\frac{{N_0 \choose (1-\nbar+\Delta \nbar) N_0}}{{N_0 \choose (1-\nbar) N_0}} x^{N_0\Delta \nbar} \exp(-\beta \Delta E_\text{pot}) \right)
\end{equation}
after inserting Eqn.~\eqref{eq:const pH probability of state} into Eqn.~\eqref{eq:tilde_acc}.
The equation above can be reformulated for a single deprotonation step in order to read
\begin{equation}
\tilde{\text{acc}}(\text{diss}|\text{ass})=\min \left(1,\frac{{N_0 \choose (1-\nbar+1/N_0) N_0}}{{N_0 \choose (1-\nbar) N_0}} x \exp(-\beta \Delta E_\text{pot}) \right)
\end{equation}
with $\Delta \nbar=1/N_0$.
By using the relation
\begin{align}
\frac{{N_0 \choose (1-\nbar+1/N_0) N_0}}{{N_0 \choose (1-\nbar) N_0}}=\frac{N_0 \nbar}{N_0 (1-\nbar) +1} \overset{N_0 \to \infty}{\sim} \frac{N_0 \nbar}{N_0 (1-\nbar) }=\frac{N_\text{HA}}{N_{\text{A}^{-}}}
\end{align}
in the thermodynamic limit for an infinite number of titrable groups $N_0$, 
we finally obtain a simple expression for the acceptance probability in the constant pH method with a symmetric proposal probability according to
\begin{equation}
\tilde{\text{acc}}(\text{diss}|\text{ass})=\min \left(1,\frac{N_\text{HA}}{N_{\text{A}^{-}}} 10^{\text{pH}_\text{in}-\text{pK}_\text{a}} \exp(-\beta \Delta E_\text{pot}) \right)
\label{eq: acc probability for symmetric proposal}
\end{equation}
which can be also derived  for a protonation reaction.
One has to notice that the so derived acceptance probability $\tilde{\text{acc}}$ in Eqn.~\eqref{eq: acc probability for symmetric proposal}  differs from the standard acceptance probability in Eqn.~\eqref{eq:acc const pH} with regard to the prefactor $N_\text{HA}/N_{\text{A}^{-}}$ which accounts for the usage of a symmetric proposal probability. 
 
\subsection{{Similarity between the reaction ensemble and the constant pH method under certain conditions}}
\label{sec:similarity}
As it was discussed in the introduction, one can either use the reaction ensemble or the constant pH method for the  simulation of weak polyelectrolytes. In this section, we demonstrate {similarity} between the constant pH method and the reaction ensemble method in the sweeping mode, as it was introduced in section \ref{sec:mode} under certain conditions.
In terms of dissociation reactions, the reaction ensemble yields the acceptance probability 
\begin{equation}
\text{acc}^{\text{RE}}(\text{diss}|\text{ass})=\min\left(1, K_{a} \frac{N_{\text{HA}}}{\text{N}_{\text{A}^-}\,c^*(\text{H}^+)}\exp(-\beta \Delta E_\text{pot} )\right)
\label{eq: RE monoprotic dissociation}
\end{equation}
which is a simplified version of Eqn.~\eqref{eq:Metropolis_RE} with the apparent reaction constant \mbox{$K_{a}=K\beta P^0$} and the currently present proton concentration $c^*(\text{H}^+)=N_{\text{H}^+}/V$ in the simulation box.
A comparison between the acceptance probability $\tilde{\text{acc}}$ in Eqn.~\eqref{eq: acc probability for symmetric proposal} for the constant pH method with a symmetric proposal probability and Eqn.~\eqref{eq: RE monoprotic dissociation} yields that both acceptance probabilities are equal if the following relation hold
\begin{equation} 
 K_a\frac{N_{\text{HA}}}{N_{\text{A}^-}\,c^*(\text{H}^+)} = K_a\frac{N_{\text{HA}}}{N_{\text{A}^-}\,c(\text{H}_\text{in}^+)}
\end{equation}
with $10^{\text{pH}_\text{in}-\text{pK}_{\text{a}}}=K_{\text{a}}/c(\text{H}_\text{in}^+)$ that can be also expressed by
\begin{equation}
\text{pK}_\text{a}-\text{pH}^* =  \text{pK}_{\text{a}}-\text{pH}_\text{in}
\label{eq:similarity acc}
\end{equation}
with the implicit and predefined $\text{pH}_\text{in}$ value as used in the constant pH method denoted by $\text{pH}_\text{in}=-\log_{10}(c(\text{H}_\text{in}^+)/(\text{mol/L}))$ including the virtual proton concentration $c(\text{H}_\text{in}^+)$ {and $\text{pH}^*=-\log_{10}(c^*(H^+)/(\text{mol/L}))$ the current pH in the simulation box}. Eqn.~\eqref{eq:similarity acc} is valid for the sweeping operational mode in the reaction ensemble method and for $\text{pH}^*=\text{pH}_\text{in}$.  If these requirements are fulfilled, the reaction ensemble method and the {reformulated} constant pH method reveal equal acceptance probabilities. {Since the current ${\text{pH}^*}$ in the reaction ensemble simulation fluctuates around the $\text{pH}:=-\log_{10}(\langle c(\text{H}^+)\rangle)$ in the reaction ensemble the average particle number in the constant pH method and the reaction ensemble in the sweeping mode are the same if $\text{pH}=\text{pH}_\text{in}$. However the variance of the particle number is typically different in both methods.}\\
{At this point, it is important to note that the pH value in the reaction ensemble method (pH) is measured via the actual proton concentration whereas the pH value in the constant pH method (pH$_\text{in}$) represents a constant input parameter.} Hence, a change of the box volume $V$ in the reaction ensemble method induces a variation of the \textit{measured} pH value in contrast to the constant pH method {where the pH is a fixed number}. It thus follows, that the reaction ensemble method allows the study of concentration dependent effects in terms of the law of dilution, which enforces a more pronounced deprotonation behavior for lower concentrations of titrable groups \cite{atkins10a}.
In fact, the constant pH method can be interpreted as a coupling scheme to an \textit{implicit} proton bath of infinite dimensions which fixes the pH value of the solution.
In contrast to the reaction ensemble method in the real titration mode, the absence of {all} free protons in the constant pH method {reduces} a screening of electrostatic interactions between the charged titrable groups. Thus, the resulting deprotonation behavior differs between the methods which can be recognized by differences in the titration curves as it will be discussed in the next sections.

\section{Simulation details}
We study the properties of weak polyelectrolytes in terms of a coarse-grained bead-spring model with $N_0=50$ beads.
{All titrable groups (beads) repel each other by a truncated and shifted Lennard-Jones
potential \cite{weeks71a} with amplitude $\epsilon = 1 k_BT$ and range $1\sigma$ yielding a cutoff radius  $r_c=2^{1/6}\sigma$. Electrostatic interactions were calculated by the P$^3$M method \cite{hockney88a} with a Bjerrum length  $\lambda_B=e^2/4\pi\varepsilon_0\varepsilon_{\rm r}k_{\rm B}T=2\sigma$ including the dielectric constant $\epsilon_{\rm r}$ and the elementary charge $e$. In comparison to an aqueous solution at room temperature yielding $\lambda_B= 0.71$ nm, we thus identify $\sigma = 0.355$ nm.
The cubic simulation box with periodic boundary conditions in all three dimensions has a box length of $b=56.3124\sigma$ with a monomer or titrable group concentration of $c_0=0.00028\sigma^{-3}$ 
and a polymer concentration of $c_p=5.6\cdot 10^{-6} \sigma^{-3}$. With the Avogadro constant $N_A=6.022\cdot 10^{23}$ mol$^{-1}$ and $\sigma=0.355$ nm, these values correspond to concentrations of $c_0\approx 0.01$ mol/L and $c_p\approx 2\cdot 10^{-3}$ mol/L.}
Bonds between adjacent beads of the polyelectrolyte are modeled by a FENE potential \cite{kremer90a} according to
\begin{equation}
U_\text{FENE}(r) = -\frac{1}{2}k r_\text{max}^2 \log\left(1-\left(\frac{r-r_0}{r_\text{max}}\right)^2\right)
\end{equation}
with the spring constant $k=10 \epsilon/\sigma^2$, the maximum elongation $r_\text{max}=1.5\sigma$ and an equilibrium length $r_0=2^{1/6} \sigma \approx 1.12\sigma$.
We perform Langevin Dynamics simulations according to
\begin{equation}
m_i\ddot{\vec{r}}_i = -\zeta\dot{\vec{r}}_i + {\vec{R}}_i + \vec{F}_i
\end{equation}
with the mass $m_i=1 m$ for each particle, the conservative force $\vec{F}_i$, the friction force $-\gamma\dot{\vec{r}}_i $ and the random force ${\vec{R}}_i$. The random force acts on each particle independently and obeys the fluctuation-dissipation theorem $\langle R_{ik}\rangle = 0$ and $ \langle R_{ik}(t) R_{jl}(t^{\prime})\rangle = 2\gamma k_BT \delta_{ij}\delta_{kl}\delta(t-t^{\prime}) $
which ensures the presence of Gaussian white noise for particles $i$ and $j$ in the spatial directions $k$ and $l$. The friction coefficient has a value of
$\gamma=1\sigma^{-1}(m\epsilon)^{1/2}$. The temperature is $T = 1 \epsilon/k_B$ and the Langevin equation is integrated by a Velocity Verlet algorithm with a time step of $\delta t = 0.01 \sigma(m/\epsilon)^{1/2}$. \\
The apparent reaction constant $K_a=K\beta P^0$ in the reaction ensemble depends on the standard pressure $P^0=1\text{ bar} =0.00108 \epsilon/\sigma^3$  and on $\beta = 1\epsilon^{-1}$. 
For the constant pH method, we varied the values for $\rm{pK_a-pH}$ between -4 and 2. In the sweeping mode of the reaction ensemble, we choose dimensionless reaction constants $K$ between the values -8 and 2 in logarithmically equidistant intervals. In contrast, we consider values of $\text{pK}_a=0.49$ and $\text{pK}_a=3$ in the real titration mode and vary  the number of negatively and positively charged excess protons or hydroxide ions at specific pH values. Moreover, we also take the autoprotolysis of water into account by adding the apparent autoprotolysis reaction constant $K_\text{w}=10^{-14} \rm{mol^2/L^2}=10^{-14} \cdot (0.02694/\sigma^3)^2$ \cite{atkins10a}. 
All simulations are performed  with the software package {\sf ESPResSo}~\cite{espresso1,espresso2}. 

\FloatBarrier
\section{Numerical Results}
\begin{figure}
\centering
\includegraphics[width=\linewidth]{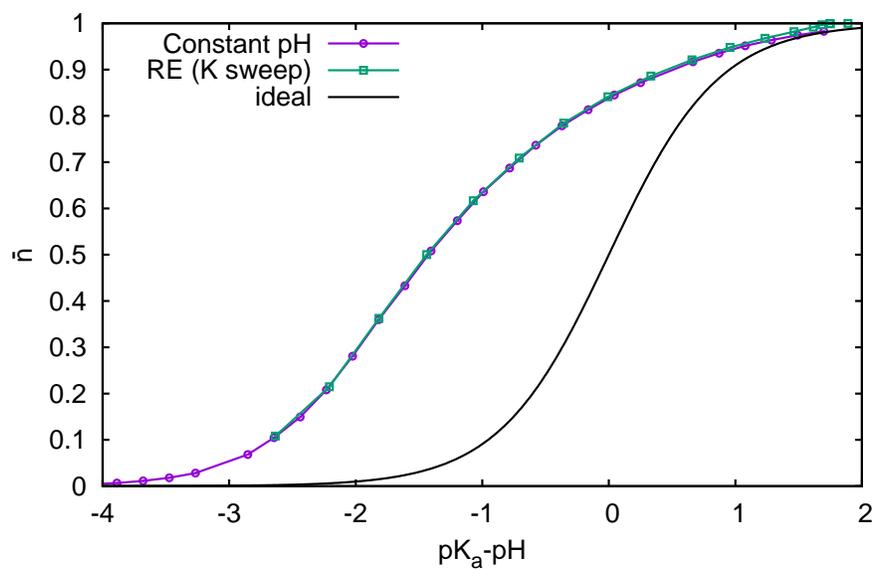}
\caption{Titration curve for a flexible polyelectrolyte with Bjerrum length $\lambda_B=2\sigma$ as simulated by the constant pH method and the reaction ensemble method in the sweeping mode. The black solid line corresponds to an ideal titration curve without conservative interactions between the charged groups and particles in the system.  
}
\label{fig:reaction_ensemble_vs_const_pH}
\end{figure}
\begin{figure}
\centering
\includegraphics[width=\linewidth]{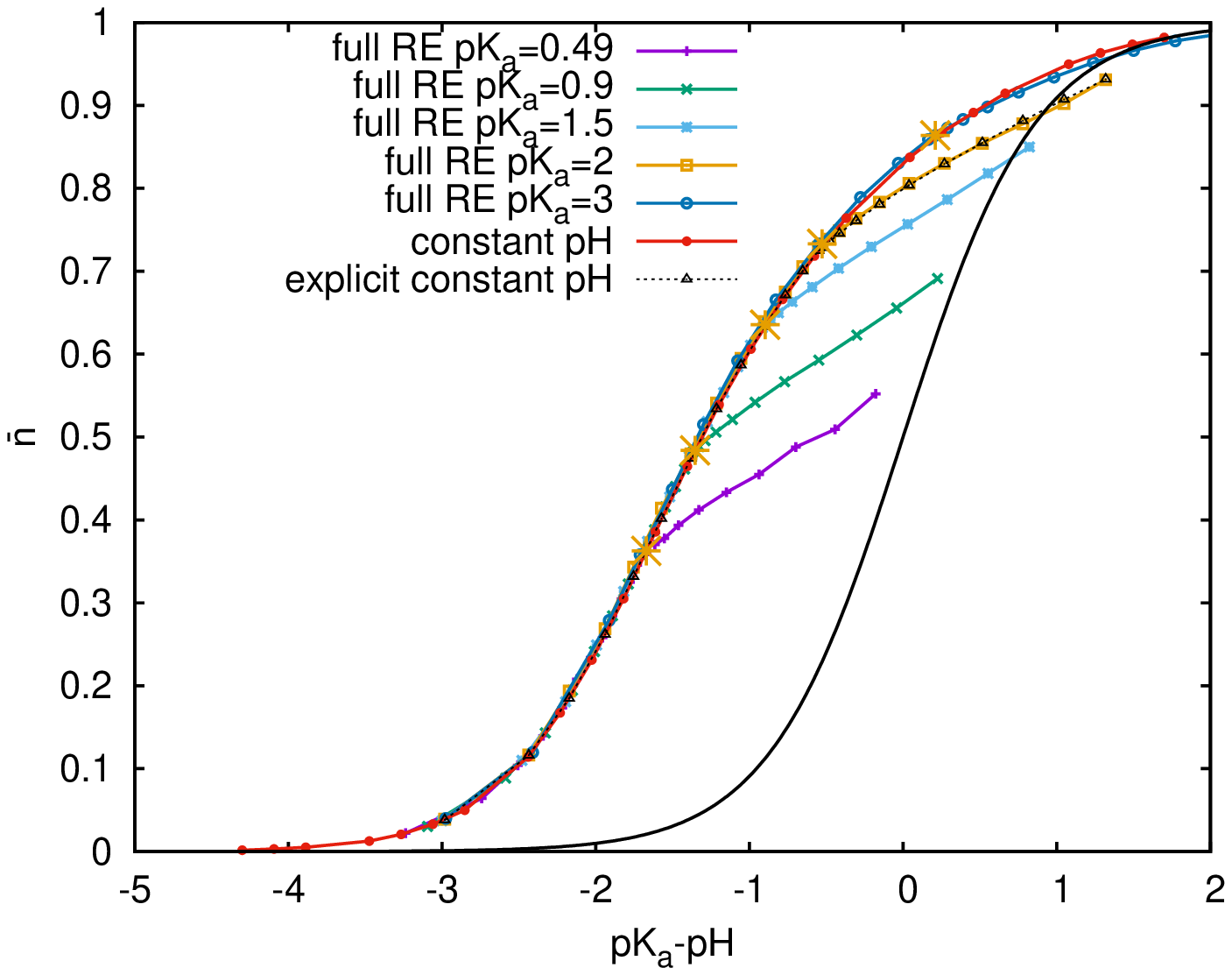}
\caption{Titration curves for flexible weak polyelectrolytes with different pK$_a$ values and a Bjerrum length $\lambda_B=2\sigma$ as obtained by the constant pH method and the reaction ensemble method in the real titration mode. {The orange stars denote the pK$_\text{a}-\text{pH}$ values where a strong acid was injected in order to reach lower pH values according to the point of the eigen pH value.}
The black solid line represents an ideal titration curve without conservative interactions. 
{The dashed black line represents the results of a modified constant pH method for pK$_\text{a}=2$ in order to study electrostatic screening effects. More information can be found in the main text.} 
}
\label{fig:real_titration_reaction_ensemble_vs_const_pH}
\end{figure}
\begin{figure}
\centering
\includegraphics[width=\linewidth]{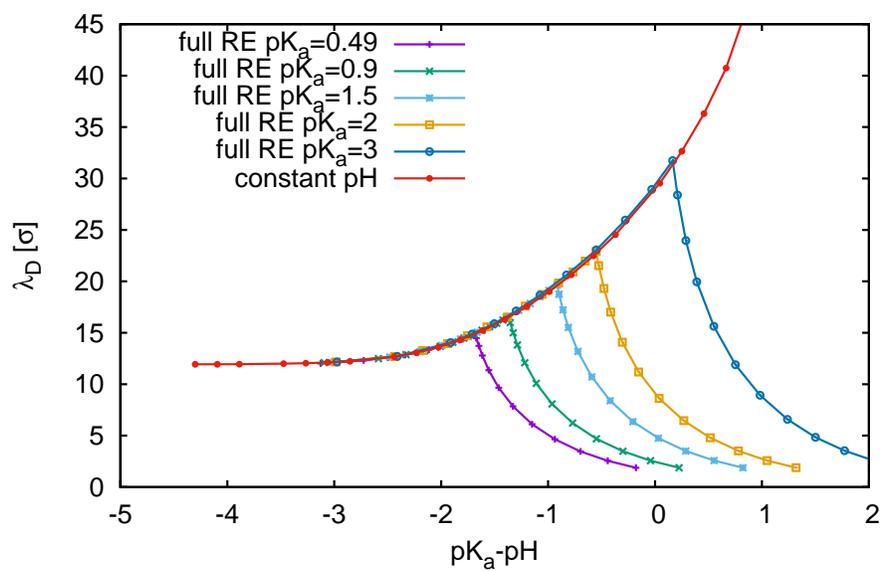}
\caption{{Electrostatic Debye-H{\"u}ckel screening lengths for flexible weak polyelectrolytes with different pK$_a$ values and a Bjerrum length $\lambda_B=2\sigma$ as obtained by the reaction ensemble method in the real titration mode shown in Fig.~\ref{fig:real_titration_reaction_ensemble_vs_const_pH}. and the constant pH method.}}
\label{fig:screening}
\end{figure}

In Fig.~\ref{fig:reaction_ensemble_vs_const_pH}, we compare the titration curves between the constant pH and the reaction ensemble method in the sweeping mode for a flexible weak polyelectrolyte with Bjerrum length $\lambda_B=2\sigma$. As it was discussed in Sec.~\ref{sec:similarity}, a nearly perfect agreement between the results {for the reaction ensemble and the constant pH method} can be observed. 
Moreover, strong deviations to an ideal titration curve with $\Delta E_{\text{pot}} = 0$  can be seen. \\
The results of the reaction ensemble method in the real titration mode and the constant pH method for different pK$_a$ values are presented in  Fig.~\ref{fig:real_titration_reaction_ensemble_vs_const_pH}.
Depending on the intrinsic $\text{pK}_a$ value of the titrable units and the pH value of the solution, the titration curves of the reaction ensemble deviate significantly from those obtained by the constant pH method.  As an example, for a moderately strong acid with $\rm{pK_a=0.49}$, one can observe pronounced differences between both curves at $\text{pK}_a-\text{pH} \geq -1.67$ corresponding to low pH values.  
{The differences to the constant pH method can be mainly attributed  to the additionally occurring electrostatic screening effects in the real titration mode. In order to achieve higher association degrees \nbar and lower pH values, a strong acid is injected into the system whose chemical species \ce{H+} and \ce{Cl-} induce a screening of electrostatic interactions between the deprotonated titrable groups \cite{hunter01a}. The corresponding points of eigen pH values for the given choice of the reaction constant, as introduced in section \ref{sec:mode}, are denoted by orange stars in Fig.~\ref{fig:real_titration_reaction_ensemble_vs_const_pH}.
Hence, for very low pH values and high concentrations of the excess strong acid, the charges of the titrable units are screened such that the polyelectrolyte becomes more and more ideal for decreasing pH values.  Furthermore, it  can be observed that for polyelectrolytes with pK$_a \leq 2$, the degrees of association for $\text{pK}_a-\text{pH}\geq 0.25$ are significantly smaller compared with the ideal titration curve.
We can attribute this finding to the explicit presence of  charged species in the solution. It was experimentally found and in depth discussed in Refs.~\cite{ramos99a,panagiotopoulos2009charge} that the presence of ions, like in a salty solution favors a stronger dissociation of polyelectrolytes. Thus, the charged species stemming from the injected strong acid resemble a salty solution such that lower degrees of association at specific pH values in comparison to ideal titration curves can be observed.}\\
{In order to verify the presence of electrostatic screening effects, we studied a polyelectrolyte with pK$_\text{a}=2$ in presence of explicit salt ions and by using the constant pH method (gray triangles in Fig.~\ref{fig:real_titration_reaction_ensemble_vs_const_pH}). More specifically, we calculated the Debye H{\"u}ckel length $\lambda_D=\sqrt{1/(4\pi \lambda_B \sum_i c_i z_i^2)}$ \cite{hunter01a} where $c_i$ is the concentration of charged species with valency $z_i$ in the reaction ensemble in the real titration mode at distinct pH values (yellow squares) and added the corresponding concentrations of chemically inert salt anions and cations to the constant pH simulations. The coincidence between the curves verifies our assumption that electrostatic screening effects are mainly responsible for the differences between the constant pH and the reaction ensemble method.}\\
{
Moreover, due to the increasing number of protons from the injected strong acid at pH values lower than the point of the eigen pH value, a significant decrease of the electrostatic Debye-H{\"u}ckel screening length can be observed at the right side of Fig.~\ref{fig:screening}.
For low pH values, the results for the constant pH method indicate that the Debye-H{\"u}ckel length diverges in comparison to the reaction ensemble method in the real titration mode, which can be attributed to the above discussed absence of explicit excess free protons in the constant pH method. Moreover, after a comparison between Fig.~\ref{fig:real_titration_reaction_ensemble_vs_const_pH} and Fig.~\ref{fig:screening}, one can observe that most pronounced differences between the methods are evident for $\lambda_D\leq 10\sigma$ which is in the order of the polyelectrolyte size. 
{\em Vice versa}, in order to simulate higher pH values, a strong base is added to the system whose hydroxide ions annihilate with the free protons of the weak polyelectrolyte in terms of autoprotolysis reactions (left side of  Fig.~\ref{fig:screening}).  Thus, the Debye-H{\"u}ckel lengths and the titration curves are identical in the constant pH and the reaction ensemble method for high pH values until the point of the eigen pH value is reached. The Debye-H{\"u}ckel length also decreases for higher pH values, due to a significantly more pronounced dissociation of the weak acid aresulting in a high amount of  free charged species. 
Due to these reasons, we conclude that electrostatic screening effect impose a significant influence on weak polyelectrolytes which complicates the applicability of the constant pH method at low pH values. However for weak polyelectrolytes which have a high enough $\text{pK}_a$ value (e.g. $\text{pK}_a=3$), the difference between the constant pH titration curve and the reaction ensemble titration curve in the real titration mode practically vanish (see Fig.~\ref{fig:real_titration_reaction_ensemble_vs_const_pH}).} \\
In summary, pH dependent screening effects are not adequately reproduced by the standard constant pH method or the reaction ensemble in the sweeping mode due to the fact that both approaches consider the pH value implicitly. Nevertheless, it can be assumed that the explicit treatment of pH dependent screening effects might be relevant for proteins, based on the findings that the individual amino acids strongly differ in their deprotonation/protonation properties \cite{berg02a}.

\FloatBarrier
\section{Summary and conclusion}
In this article, we demonstrated the similarity between the reaction ensemble method in the sweeping mode and the constant pH method {under certain conditions}. Both methods can be used to study the dissociation properties of titrable groups in weak polyelectrolytes. Noteworthy, the implicit interpretation of the pH value in the constant pH method inhibits electrostatic screening effects due to the absence of all explicit free protons according to the pH value of the solution around the charged groups of the polyelectrolyte. This finding points at certain complications of the constant pH method.
{It is evident that these effects are mostly important at extremely high or low pH values. In fact, for moderate pH values, the constant pH and the reaction ensemble method reveal comparable results.}
Moreover, we proposed a new operational mode for the reaction ensemble method, which can be used to study the behavior of polyelectrolytes according to real titration procedures. Based on our findings, we conclude that the usage of the reaction ensemble method in the real titration mode is specifically preferred for polyelectrolytes with different functional groups, under confinement and for the simulation of acids and bases with moderate pK$_a$ values.  

\section{Acknowledgments}
We thank Tobias Richter, Peter Kosovan and Kai Szuttor for fruitful discussions and helpful ideas. Financial funding by the Deutsche Forschungsgemeinschaft  through the grants AR 593/7-1 and {HO 1108/26-1} is gratefully acknowledged. \\
{We dedicate this work to Wolfhard Janke on the occasion of his 60th birthday and wish him an equally scientific fruitful future.}

\bibliographystyle{mystyle}
\bibliography{literature}

\end{document}